\documentclass{aastex631}

\usepackage{rotating}
\usepackage{amsmath}
\usepackage{soul}
\usepackage{lipsum}
\usepackage{amsthm,amsmath,amssymb}

\hypersetup{linkcolor=red,citecolor=blue,filecolor=cyan,urlcolor=magenta}

\begin{document}

\title{Detailed Investigation of a W UMa Contact Binary with an Ultralow Mass Ratio and a Third-Body as a Potential Merger Candidate}

\author[0000-0002-0196-9732]{Atila Poro}
\altaffiliation{atilaporo@bsnp.info, atila.poro@obspm.fr}
\affiliation{Binary Systems of South and North (BSN) Project, 15875 Tehran, Iran}
\affiliation{LUX, Observatoire de Paris, CNRS, PSL, 61 Avenue de l'Observatoire, 75014 Paris, France}

\author[0000-0001-9746-2284]{Ehsan Paki}
\affiliation{Binary Systems of South and North (BSN) Project, 15875 Tehran, Iran}

\author[0000-0002-1972-8400]{Fahri Alicavus}
\affiliation{Çanakkale Onsekiz Mart University, Faculty of Science, Department of Physics, 17020, Çanakkale, Türkiye}
\affiliation{Çanakkale Onsekiz Mart University, Astrophysics Research Center and Ulupnar Observatory, 17020, Çanakkale,
Türkiye}

\author[0000-0003-1263-808X]{Raul Michel}
\affiliation{Instituto de Astronom\'ia, UNAM. A.P. 106, 22800 Ensenada, BC, M\'exico}

\begin{abstract}
The lower limit of the mass ratio in contact binaries remains uncertain, with observations suggesting systems exist below theoretical predictions. The stability of such very low mass ratio systems is still debated. Based on our review of systems within the mass ratio cutoff range, we reanalyzed TYC 3801-1529-1 and found it to have the lowest known mass ratio, $q = 0.024_{-(1)}^{+(2)}$, among analyzed contact binaries. The reanalysis of this target was carried out using the BSN application and the MCMC method. We then compared our light curve solution obtained from TESS observations with the results of a previous study. We studied the period variations of this system and identified a cyclic trend over the past six years. After the third-body contribution has been removed, the system's period variations can be described either by a linear trend with a negative slope or by a quadratic trend with a downward curvature. These results indicate that it is still not possible to definitively determine whether the orbital period is increasing or decreasing, underscoring the importance of future observations. By considering the challenges of detecting the faint secondary in extremely low mass ratio systems, we estimated the absolute parameters of the target. Based on our analysis, the secondary component of the binary is likely a brown dwarf, while the third body appears to be a low-mass M-type dwarf. According to our analysis, TYC 3801-1529-1 is dynamically unstable and thus represents a remarkable candidate for a binary merger.
\end{abstract}


\section{Introduction}
W UMa contact binaries are short-period systems in which both stars fill their Roche lobes and are embedded in a shared convective envelope. This common envelope redistributes the energy generated by the components, leading to nearly identical surface temperatures (\citealt{Lucy1968}). The mass ratio, defined as $q = M_{2}/M_{1}$, is a key parameter in characterizing contact binaries. The mass ratio in contact binary systems can be determined with high accuracy when both photometric and spectroscopic data are available. Nevertheless, photometric observations of systems with high orbital inclinations and total eclipses can also provide reliable estimates of the mass ratio (\citealt{2021AJ....162...13L, 2024AJ....168..272P}). It should be noted, however, that in systems with extremely low mass ratios, detecting the secondary component in spectroscopic data can be particularly challenging.

Most studies of contact binaries with extremely low mass ratios have relied primarily on photometric modeling, either alone or combined with spectroscopic observations in which the secondary component is not detected. The principal difficulty in obtaining a spectroscopic mass ratio in these systems is the intrinsically low luminosity of the secondary star. When the mass ratio is extremely low, the secondary contributes only a minor fraction of the system's total light, resulting in absorption features that are shallow and easily obscured within the broader and deeper lines of the primary (\citealt{2025MNRAS.540.1290L}). Additional complications arise from rapid rotation and tidal distortion, which produce significant line broadening in both components; when the secondary is faint, this broadening effectively merges its spectrum into that of the primary (\citealt{2008MNRAS.386..377P}). Achieving a reliable detection of the secondary component in such systems requires high-resolution spectra, typically with resolving powers of $R \sim 20{,}000$–$50{,}000$, and instrumental setups suitable for short-period, rapidly rotating systems. Careful line-profile disentangling or cross-correlation techniques are also necessary to recover the weak signatures of the secondary component (\citealt{2015AJ....149...49R}). The exact observational requirements depend on the system's brightness, orbital period, rotational broadening, and the capabilities of the telescope and spectrograph used. Consequently, several reported systems with extremely low mass ratios, such as VSX J082700.8+462850 (\citealt{2021ApJ...922..122L}) and TYC 4002-2628-1 (\citealt{2022MNRAS.517.1928G}), are based solely on photometric solutions, where degeneracies among inclination, fillout factor, and potential third light may influence the inferred mass ratio. Systems for which precise radial velocity measurements are available provide substantially more robust constraints. A well-known example is SX Crv, which has a spectroscopically confirmed mass ratio of $q = 0.066 \pm 0.003$ (\citealt{2001AJ....122.1007R}).

The dynamical stability or instability of systems with extremely low mass ratios (generally $q < 0.1$) remains open to discussion. Contact binary systems with extremely low mass ratios, which are close to the theoretical minimum for dynamical stability and serve as potential progenitors of mergers, have attracted significant attention in the literature. Theoretical and statistical studies have progressively refined estimates of the minimum mass ratio, with early calculations by \cite{1995ApJ...444L..41R} suggesting $q_{\rm min} \sim 0.09$, while subsequent work considering the angular momentum of both components found lower limits around 0.076-0.078 (\citealt{2006MNRAS.369.2001L}). Later models, incorporating the structural properties of both the primary and secondary stars, further reduced the predicted minimum mass ratio to a range of 0.05-0.109 (\citealt{2007MNRAS.377.1635A,2009MNRAS.394..501A,2010MNRAS.405.2485J}). Statistical analyses of overcontact, low mass ratio systems indicated that the minimum ratio could be as low as 0.044 (\citealt{2015AJ....150...69Y}), and more recent studies suggest a range of 0.038-0.041 depending on the primary mass, stellar structural parameters, and metallicity (\citealt{2021MNRAS.501..229W,2024NatSR..1413011Z,2024SerAJ.208....1A,2024MNRAS.527....1W}). Despite these predictions, only one confirmed contact binary merger, V1309 Sco, has been observed and verified to date (\citealt{2011A&A...528A.114T}), highlighting the rarity and astrophysical significance of such systems.

A notable example of extremely low mass ratio contact binaries is VSX J082700.8+462850, studied by \cite{2021ApJ...922..122L}, which has a reported mass ratio of 0.0550(5). Subsequently, TYC 4002-2628-1 was analyzed by \cite{2022MNRAS.517.1928G}, with a mass ratio of 0.0482(1). This was followed by TYC 3801-1529-1, investigated by \cite{2024A&A...692L...4L}, exhibiting a mass ratio of 0.0356(35) and an orbital period of 0.365909 days. Recently, ASASSN-V J175200.35+361805.2 was reported by \cite{2025AJ....170..101G}, with a mass ratio of $0.027(1)$, representing one of the lowest mass ratios identified in a contact binary system. Collectively, these systems highlight the existence and astrophysical significance of contact binaries near the theoretical minimum mass ratio, providing crucial observational benchmarks for studies of dynamical stability and the potential for stellar mergers.
\\
\\
We carried out a re-analysis of several extremely low mass ratio contact systems using TESS light curve data. For this purpose, we examined six systems with mass ratios below 0.09 that have been reported in recent years. Among these, the TYC 3801-1529-1 (TIC 445065429) system stood out as particularly notable, as our re-analysis yielded results that differed from those previously reported. The object TYC 3801-1529-1 was first reported as a contact binary by \cite{2020ApJS..249...18C}. The International Variable Star Index (VSX\footnote{\url{https://vsx.aavso.org/}}) and the Zwicky Transient Facility (ZTF, \citealt{2023AA...675A.195S}), both of which classify it as a contact system, provide an orbital period of 0.3659090 days, a maximum brightness of $12.537^{\textit{mag}}$, and a light variation amplitude of only $0.108^{\textit{mag}}$ in the $g$ band. Owing to this remarkably small amplitude variability, the system is regarded as a candidate for an extremely low mass ratio contact binary (\citealt{2022csss.confE.178P}). TYC 3801-1529-1 first time analyzed by \cite{2024A&A...692L...4L}. They used TESS and ground-based data along with spectroscopic observations. \cite{2024A&A...692L...4L} reported $q=0.0356(35)$ and a medium fillout factor $f=0.347(8)$ for this A-subtype W UMa contact system. Therefore, in the present study, we conducted a detailed investigation of this system, focusing on its light curve solutions, orbital period variations, and its potential for merger. The structure of this paper is as follows: Section 2 presents the light curve analysis and the estimation of the absolute parameters of TYC 3801-1529-1; Section 3 examines the orbital period variations of the target system; and Section 4 discusses the results and presents the conclusions.

\vspace{0.6cm}
\section{Light Curve Analysis}
\subsection{Photometric Solutions}
TYC 3801-1529-1 was observed by the TESS mission in sectors 20, 47, 60, and 74 with exposure lengths of 1800, 600, 200, and 200 seconds, respectively. The TESS data for this system are of good quality, and the mean photometric uncertainties across the four observed sectors are nearly identical. As the Science Processing Operations Center (SPOC, \citealt{2016SPIE.9913E..3EJ}) light curves are not available for this target, we employed the Quick-Look Pipeline (QLP, \citealt{2021RNAAS...5..234K}) products for the photometric analysis. The light curve analysis was performed using selected data from all four available TESS sectors, combined to construct the final dataset for modeling the system. The observational data were phased using the new ephemeris derived in this study (Section 3).

We first validated the results of \cite{2024A&A...692L...4L} using version 2.4.9 of the PHysics Of Eclipsing BinariEs (PHOEBE) Python code (\citealt{2016ApJS..227...29P, 2020ApJS..250...34C}) on the data under consideration. Validation through PHOEBE modeling indicates that the synthetic light curve of \cite{2024A&A...692L...4L} provides a poor fit to the observational data. The \cite{2024A&A...692L...4L} study utilized the Wilson–Devinney code (\citealt{1971ApJ...166..605W, 1979ApJ...234.1054W, 1990ApJ...356..613W}), which has slight differences compared to PHOEBE. The observed discrepancy may arise from the necessity of employing a different mesh representation for the stellar surfaces. To display and analyze this system, we set b['ntriangles']=15000 in PHOEBE. The gravity-darkening coefficients were set to $g_1 = g_2 = 0.32$ \citep{1967ZA.....65...89L} and the bolometric albedos to $A_1 = A_2 = 0.5$ \citep{1969AcA....19..245R}, typical for stars with convective envelopes. The stellar atmosphere model followed \cite{2004AA...419..725C}, with the logarithmic limb-darkening (default in PHOEBE for eclipsing binaries) adopted. The limb-darkening coefficients were treated as free parameters to achieve a more flexible and accurate light curve fit. We began our analysis using the same parameter values as reported by \cite{2024A&A...692L...4L}. The adopted effective stellar temperatures were found to be in good agreement with Gaia DR3 value (\citealt{2025MNRAS.537.3160P}), and our $q$-search indicated that the system's mass ratio is below 0.1. Moreover, as the difference between the two maxima in the light curve was below 0.003 flux and within the observational uncertainties, it was considered negligible, permitting the analysis to be conducted without introducing a starspot. Consistent with \cite{2024A&A...692L...4L}, we examined the possible contribution of a third light ($l_3$). However, in our TESS-based modeling the third-light term is negligible. In the study by \cite{2024A&A...692L...4L}, the third-light effect was more prominent in ground-based observations across different filters. After verifying that nearby stars do not introduce significant contamination, we excluded the third-light component from our final solution. We then utilized PHOEBE's optimization tool to improve the light curve solution, obtaining more accurate estimates of the effective temperatures, mass ratio, fillout factor, and orbital inclination.

\begin{figure*}
\centering
\includegraphics[width=0.49\textwidth]{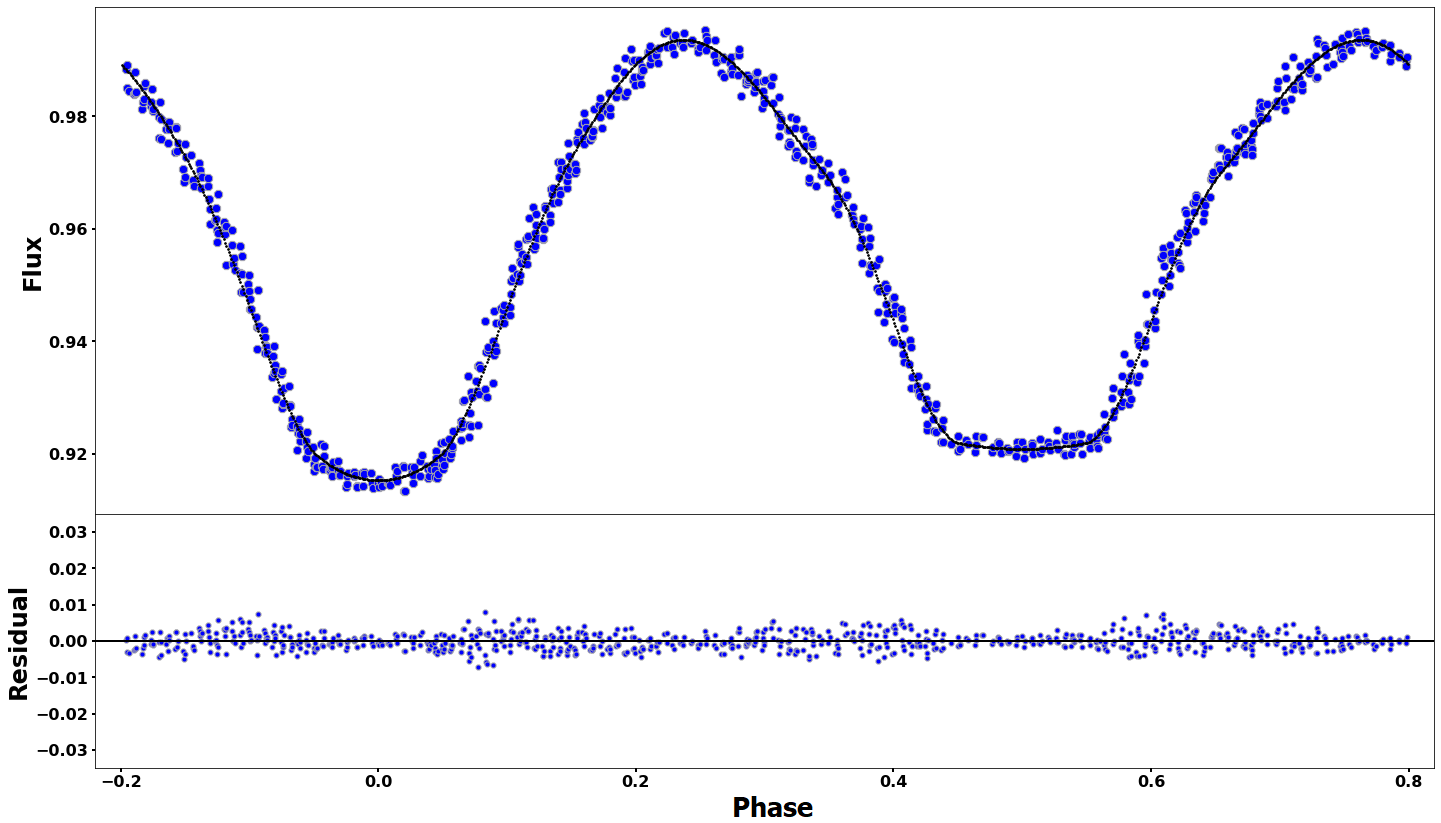}
\caption{Observed TESS light curve together with synthetic model.}
\label{fig:lc}
\end{figure*}

\begin{figure*}
\centering
\includegraphics[width=0.7\textwidth]{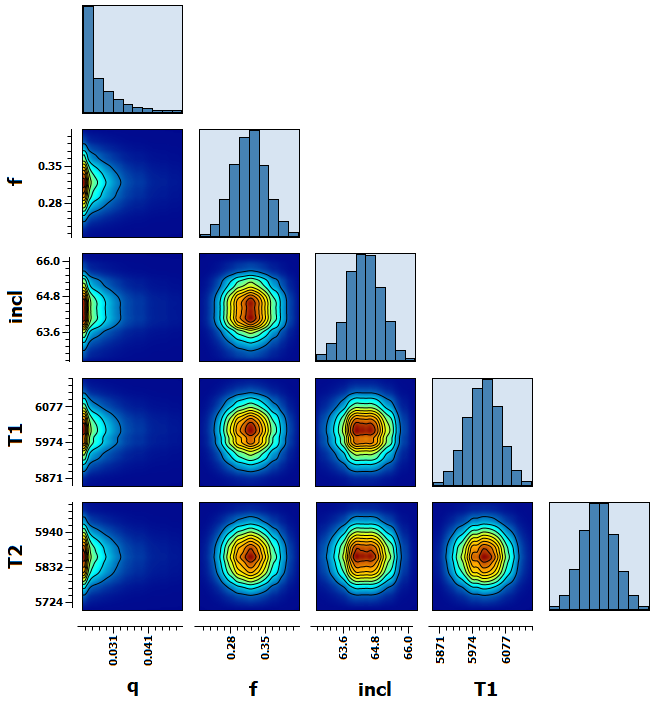}
\caption{The corner plots of TYC 3801-1529-1 were determined by MCMC modeling.}
\label{fig:corner}
\end{figure*}

\begin{figure*}
\centering
\includegraphics[width=0.8\textwidth]{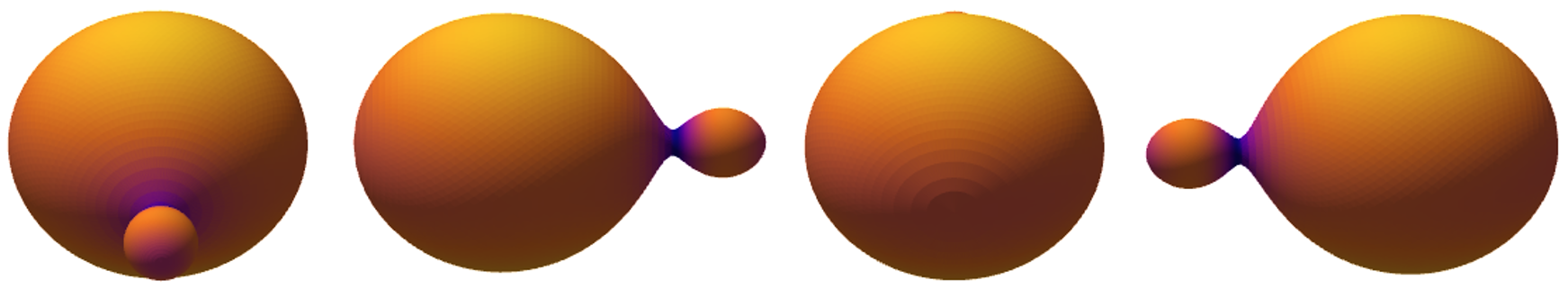}
\caption{3D views of the stars in the system at orbital phases 0, 0.25, 0.5, and 0.75, respectively.}
\label{fig:3d}
\end{figure*}

The built-in modeling and optimization routines in PHOEBE do not directly provide uncertainties for fitted parameters. In addition, the high-resolution mesh representing the stellar surfaces substantially reduces the speed of the Markov Chain Monte Carlo (MCMC) process. We employed the BSN application version 1.0 \citep{paki2025bsn}, a Windows-compatible program developed to accelerate MCMC fitting. When generating synthetic light curves, BSN achieves computational speeds more than 40 times faster than PHOEBE. The BSN application was configured with 28 walkers and 1200 iterations for the MCMC procedure, discarding the first 200 iterations as burn-in, which provided sufficient convergence while sampling the five main parameters: $T_1$, $T_2$, $q$, $f$, and $i$. The final synthetic light curve fit was identical in both the PHOEBE and BSN programs (Figure \ref{fig:lc}). Figure \ref{fig:corner} displays the corner plots, illustrating the distributions and correlations of the parameters from the MCMC analysis. The resulting parameter values along with their uncertainties are listed in Table \ref{tab:analysis}. Figure \ref{fig:3d} presents three-dimensional representations of the binary systems, constructed using the final model parameters.

\vspace{0.4cm}
\subsection{Estimation Absolute Parameters}
In the study by \cite{2024A&A...692L...4L}, the mass ratio of the totally eclipsing contact binary TYC 3801-1529-1 was determined through a combined analysis of spectroscopic and photometric data. High-resolution spectra were obtained using the BFOSC instrument on the 2.16 m telescope, from which the radial velocities of the primary component were extracted via cross-correlation with a standard star. The available data from their observations included the radial velocities of the primary star, and they were insufficient to directly determine the mass ratio from spectroscopy. In the study by \cite{2024A&A...692L...4L}, the mass ratio was inferred using the $q$-search method in combination with simultaneous modeling of TESS and ground-based light curves, taking advantage of the total eclipsing geometry to constrain the secondary component. The individual stellar masses were then calculated using the derived mass ratio together with the orbital period, resulting in $M_1 \approx 2.096~M_\odot$ and $M_2 \approx 0.075~M_\odot$, with the mass ratio $q \approx 0.0356$. Based on our analysis, the mass ratio of the system was determined to be $0.024_{\rm-(1)}^{+(2)}$. Assuming the primary mass is similar to that derived by \cite{2024A&A...692L...4L}, the secondary mass was then calculated. Then, the semi-major axis ($a$) of the system was calculated using Kepler's third law and the values of the stellar masses and orbital period. The stellar radii were calculated using the mean relative radii ($r$) and $a$ according to the relation $R = r \times a$. Finally, the stellar luminosities were determined using the Stefan-Boltzmann law, based on the measured radii and effective temperatures. The results of the absolute parameter determinations are presented in Table \ref{tab:analysis}.

\begin{table}
\renewcommand\arraystretch{1.5}
\caption{Light curve solution and absolute parameters of TYC 3801-1529-1 on TESS photometric data.}
\centering
\begin{center}
\footnotesize
\begin{tabular}{c c c}
\hline
Parameter & This study & Li et al. (2024)\\
\hline
$T_{1}$ (K) & $6009_{\rm-(53)}^{+(51)}$ & 6206\\
$T_{2}$ (K) & $5871_{\rm-(55)}^{+(54)}$ & 6160(241)\\
$q$ & $0.024_{\rm-(1)}^{+(2)}$ & 0.036(4)\\
$f$ & $0.315_{\rm-(34)}^{+(33)}$ & 0.347(8)\\
$i^{\circ}$ & $64.31_{\rm-(58)}^{+(61)}$ & 66.7(1.8)\\
$\Omega_1=\Omega_2$ & 1.670(8) & 1.724(2)\\
$l_1/l_{tot}(TESS)$ & 0.964(6) & 0.948(17)\\
$l_2/l_{tot}(TESS)$ & 0.036(1) & 0.043\\
$l_3/l_{tot}(TESS)$ &  & 0.009(3)\\
$r_{1(mean)}$ & 0.680(7) & \\
$r_{2(mean)}$ & 0.137(8) & \\
\hline
$M_1 (M_\odot)$ & 2.096(642)* & 2.096(642)\\
$M_2 (M_\odot)$ & 0.050(16) & 0.075(30)\\
$R_1 (R_\odot)$ & 1.877(188) & 1.831(207)\\
$R_2 (R_\odot)$ & 0.378(44) & 0.434(81)\\
$L_1 (L_\odot)$ & 4.134(835) & 4.455(1.045)\\
$L_2 (L_\odot)$ & 0.153(36) & 0.243(129)\\
$a(R_\odot)$ & 2.760(275) & 2.789(282)\\
\hline
\end{tabular}
\end{center}
\label{tab:analysis}
\end{table}

\vspace{0.6cm}
\section{Orbital Period Variation}
Phasing the light curve with the updated ephemeris of \cite{2024A&A...692L...4L} yields a significant phase shift on the TESS sectors data, implying that the updated ephemeris may not provide the expected level of accuracy. Therefore, we carried out an analysis of the orbital period variations of TYC 3801-1529-1.

Ground-based photometric observations of the binary system TYC 3801-1529-1 were obtained at the San Pedro Mártir (SPM) Observatory in México, located at $115^\circ27^{'}49^{''}$ W, $31^\circ02^{'}39^{''}$ N, at an elevation of 2830 m above sea level, with the aim of determining new times of minimum light. The observations were carried out on on 20 and 23 October 2025 using the 0.84 m Ritchey–Chrétien telescope ($f/15$) at SPM, equipped with a Marconi-5 CCD (e2v CCD231-42, $15\times15,\mu$m pixels, gain $2.2\ e^-/\mathrm{ADU}$, readout noise $3.6\ e^-$). Standard $BVR_cI_c$ filters were used. Exposure times were 40 s in $B$, 20 s in $V$, 15 s in $R_c$, and 10 s in $I_c$. The images were reduced with IRAF through the usual bias and flat-field corrections following established procedures. Therefore, two times of minimum (primary and secondary) were derived from these observations using Gaussian fitting.

We extracted 408 times of minima from four TESS sectors. Additionally, one ZTF minimum and 17 ground-based mid-eclipse times reported by \cite{2024A&A...692L...4L} were collected (Table \ref{tab:minima}). These minima cover a time span of about six years. In contrast to \cite{2024A&A...692L...4L}, we did not use ground-based Wide Angle Search for Planets (SuperWASP, \citealt{2010A&A...520L..10B}) observations due to the large scatter in the data. The extracted and utilized times of minima in this study are available in an online machine-readable format.

\begin{figure*}
\centering
\includegraphics[width=0.90\textwidth]{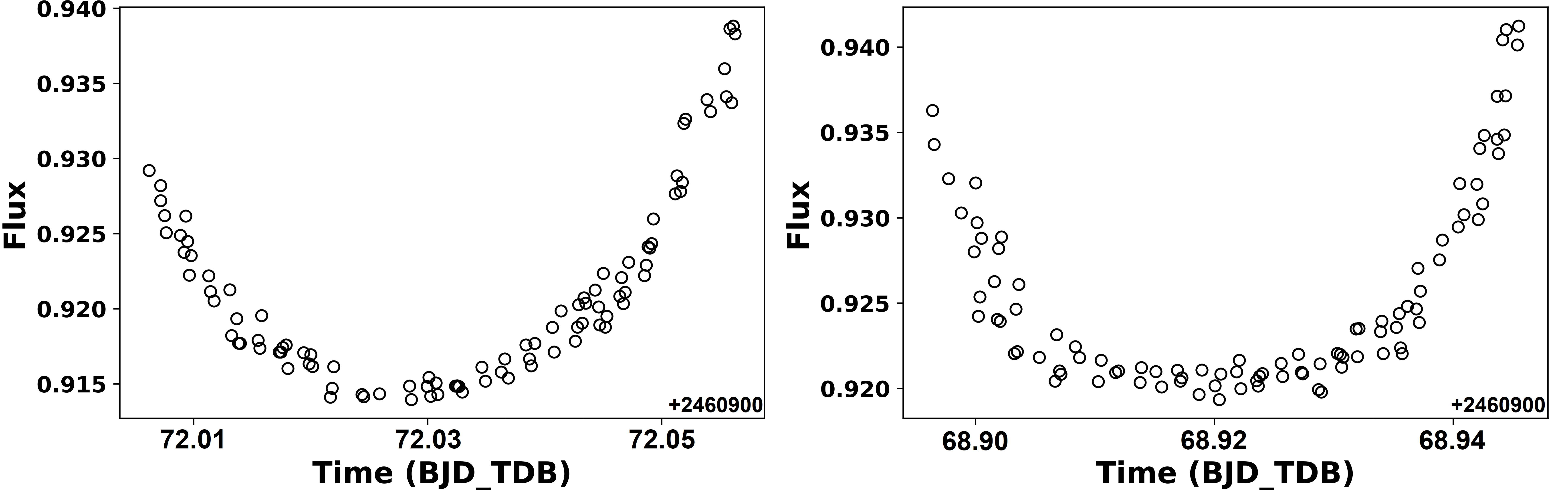}
\caption{Time of primary minimum is shown on the left, time of secondary minimum on the right, as observed in this study through the $V$ filter.}
\label{fig:min}
\end{figure*}

\begin{table*}
\centering
\caption{Observed light minima and corresponding O–C values of the target system based on data from our observations, TESS, ZTF, and \cite{2024A&A...692L...4L}. The complete table in machine-readable format is available online.}
\label{tab:minima}
\begin{tabular}{lccccc}
\hline
Min.($\text{BJD}_{TDB}$) & Error & Epoch & O-C & Reference\\
\hline
2458594.6791	&		&	-1808	&	0.0023	&	ZTF	\\
2458842.7782	&	0.0013	&	-1130	&	0.0078	&	TESS	\\
2459256.2597	&	0.0004	&	0	&	0	&	Li et al. (2024)	\\
2459580.4532	&	0.0007	&	886	&	-0.0114	&	TESS	\\
2459580.6358	&	0.0009	&	1860.5	&	-0.0117	&	TESS	\\
2459937.2287	&	0.0010	&	1861	&	-0.0075	&	TESS	\\
2460283.3910	&	0.0003	&	2807	&	-0.0053	&	Li et al. (2024)	\\
2460316.6848	&	0.0010	&	2898	&	-0.0102	&	TESS	\\
2460378.1593	&	0.0006	&	3066	&	-0.0103	&	Li et al. (2024)	\\
2460968.9211 & 0.0019 & 4680.5 & -0.0258 & This study\\
2460972.0321 & 0.0017 & 4689	& -0.0252 & This study\\
\hline
\end{tabular}
\end{table*}

The orbital period variations of this target were investigated using the observed minus calculated eclipse timing (O-C) method. Using a reference ephemeris similar to that employed by \cite{2024A&A...692L...4L} (Equation \ref{eq:OC-Ref}), we calculated the epochs and the corresponding O-C values.

\begin{equation}\label{eq:OC-Ref}
\text{BJD}_{TDB}=2459256.25969 + 0.36591971 \times E.
\end{equation}

The updated linear ephemeris was obtained by applying the linear terms derived from the O–C analysis to the reference ephemeris (Equation \ref{eq:OC-New}).

\begin{equation}\label{eq:OC-New}
\text{BJD}_{TDB}=2459256.25802(7) + 0.36591569(3) \times E.
\end{equation}

\begin{figure*}
\centering
\includegraphics[width=0.75\textwidth]{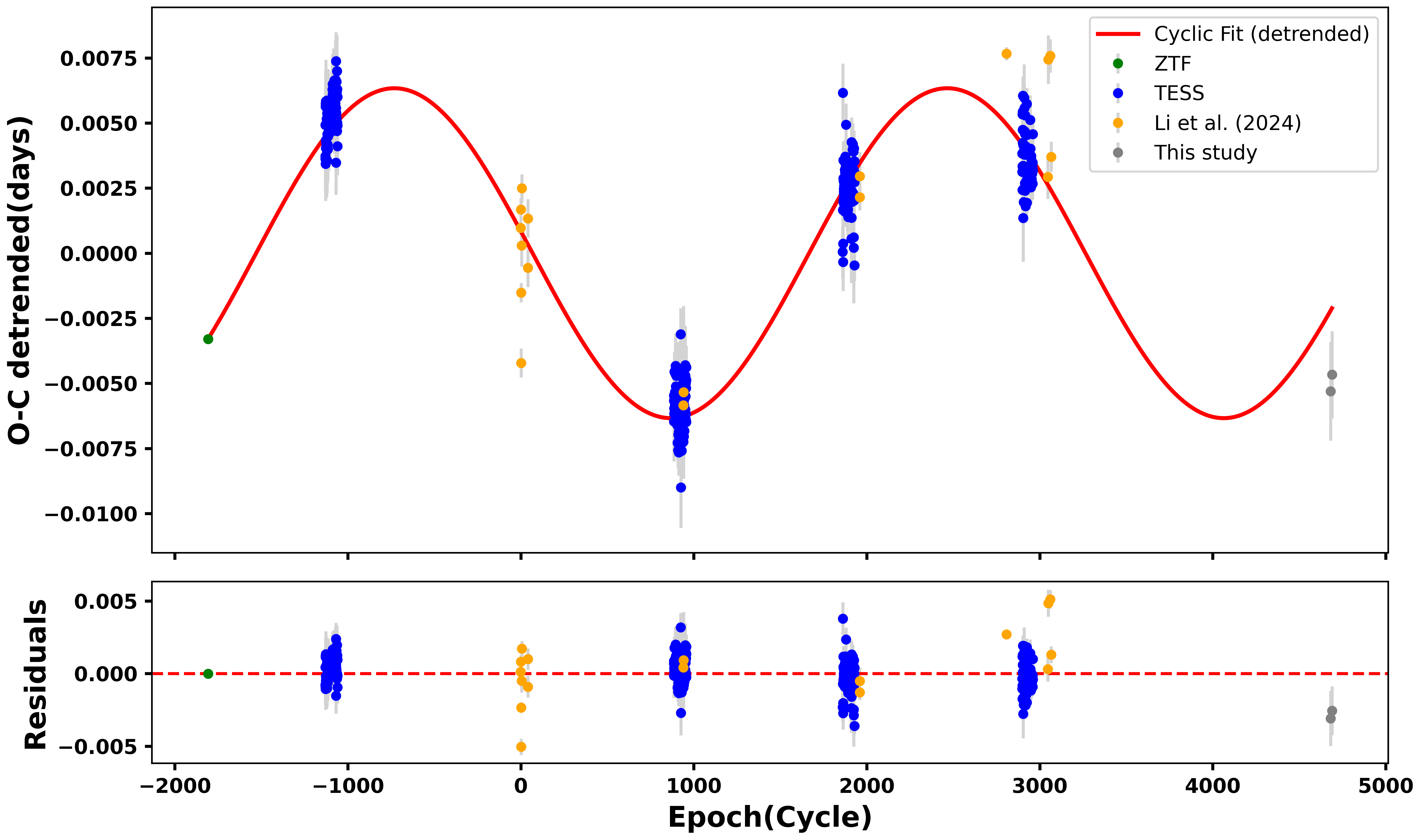}
\caption{The O-C diagram after removing the linear trend, along with its residuals. Points represent observed minima from TESS (blue), ZTF (green), the literature (orange), and the minima derived in this study (gray). The red line shows the best-fit sinusoidal model.}
\label{fig:o-c}
\end{figure*}

The O-C diagram of TYC 3801-1529-1, together with its residuals, is presented in Figure \ref{fig:o-c}. The target system clearly exhibits a cyclic variation in the orbital period. In the observed O-C diagram, the apparent linear trend may reflect slight deviations in the adopted reference ephemeris. Such a trend does not represent any intrinsic physical variation of the system but rather reflects a systematic offset between the assumed and the true linear ephemeris (\citealt{2024NewA..10502112S}). To isolate the genuine long-term or cyclic variations (e.g., due to mass transfer, magnetic activity cycles, or the light-travel time effect of a third-body), we subtracted the best-fit linear component from the data and thereby detrended the O-C curve. This process ensures that the residual O-C curve displays only the physical modulations of interest, free from artifacts associated with a referenced ephemeris.

To model cyclic behavior of the system, we used a sinusoidal O-C function:

\begin{equation}\label{eq:OC-Sin}
(O-C) = \Delta T_0 + \Delta P_0 \, E + A \sin\left(\frac{2\pi E}{P_3} + \varphi \right),
\end{equation}

\noindent where $E$ is the cycle number, $\Delta T_0$ and $\Delta P_0$ are linear corrections to the reference ephemeris, and $A$, $P_3$, and $\varphi$ describe the amplitude, modulation period, and initial phase of the sinusoidal variation. The best-fit parameters are provided in Table \ref{tab:sin3}.

To assess whether the observed cyclic changes could arise from magnetic activity, the variation in the quadrupole moment ($\Delta Q$) was estimated following (\citealt{1992ApJ...385..621A}):

\begin{equation}
\frac{\Delta P}{P} = -9 \left(\frac{R}{a}\right)^2 \frac{\Delta Q}{M R^2},
\end{equation}

\noindent where $M$ and $R$ are the mass and radius of the active stellar component, and $a$ is the semi-major axis of the binary (Table \ref{tab:analysis}). The relative period change, $\Delta P/P$, was inferred from the amplitude of the sinusoidal O-C variation.

The quadrupole moment variations for the two components were calculated using the parameters listed in Table \ref{tab:analysis} and the sinusoidal fit from Table \ref{tab:sin3}. The resulting values are approximately $1.90 \times 10^{49}$ g\,cm$^2$ for the primary and $8.33 \times 10^{49}$ g\,cm$^2$ for the secondary. These magnitudes are more than two orders of magnitude smaller than the typical quadrupole moment variations observed in close binaries, which generally range from $10^{51}$ to $10^{52}$ g\,cm$^2$ (\citealt{1999A&A...349..887L}). This substantial difference clearly demonstrates that the Applegate mechanism alone cannot account for the observed amplitude of the O-C modulation.

Consequently, the periodic O-C variations are most likely caused by a tertiary companion. The mass function of the third component was calculated using the standard relation:

\begin{equation}
f(M_{3}) = \frac{(M_{3}\sin i_{3})^{3} }{(M_{1}+M_{2}+M_{3})^{2} } = \frac{4\pi^{2}}{G P_3^{2}} (a_{12}\sin i_{3})^3,
\end{equation}

\noindent where $M_1$ and $M_2$ are the masses of the binary components, $P_3$ is the modulation period, and $a_{12}\sin i_3$ is the projected semi-major axis of the binary's orbit around the center of mass of the triple system. The resulting values are summarized in Table \ref{tab:sin3}, including Monte Carlo estimates of $M_3$ and its 1$\sigma$ confidence interval.

These findings demonstrate that the observed periodic modulation in TYC 3801-1529-1 cannot be explained solely by internal magnetic effects, and the presence of a third-body is consistent with the characteristics of the O-C variations.

\begin{table*}
\centering
\caption{Sinusoidal fit parameters and derived third-body properties for TYC 3801-1529-1. Uncertainties correspond to $1\sigma$ errors.}
\label{tab:sin3}
\begin{tabular}{lcc}
\hline
Parameter & Value & Unit\\
\hline
Sinusoidal Fit:\\
\hline
$a$ & $-0.00167 \pm 0.00007$ & days \\
$b$ & $-4.02\times10^{-6} \pm 3.12\times10^{-8}$
& days/cycle \\
$A$ & $-0.00633 \pm 0.00005$ & days \\
$P_{\rm mod}$ & $3196.07 \pm 7.80$ & cycles \\
$\phi$ & $-0.134 \pm 0.012$ & rad \\
\hline
Derived third-body Parameters:\\
\hline
$a_{12} \sin i_3$ & $1.097 \pm 0.009$ & AU \\
$f(M_3)$ & $0.0173 \pm 0.0004$ & $M_\odot$ \\
$M_{3,{\rm min}}$ & $0.493$ & $M_\odot$ \\
$M_3$ (MC mean / median / std) & $0.488 / 0.493 / 0.094$ & $M_\odot$ \\
$M_3$ (1$\sigma$ interval) & $[0.397, 0.580]$ & $M_\odot$ \\
$P_{\rm mod}$ & $8.750 \pm 0.021$ & yr \\
\hline
\end{tabular}
\end{table*}

\vspace{0.6cm}
\section{Discussion and Conclusion}
This study presents an analysis of the light curve and orbital period variations of the W UMa-type contact binary system TYC 3801-1529-1. The discussion and conclusions are as follows:

A) The mass ratio cutoff for contact binary systems has long been a challenging and debated topic in both theoretical and observational studies. We conducted a detailed light curve analysis of the system TYC 3801-1529-1, revealing a mass ratio of 0.024, which represents the lowest experimentally determined value reported to date for contact binaries.

The study by \cite{2024A&A...692L...4L} reported a mass ratio of 0.0356 with an uncertainty of 0.0035. Photometric and spectroscopic observations of this system were carried out by \cite{2024A&A...692L...4L}; however, the spectroscopic data were obtained only for the primary star. The spectral lines of the secondary are typically very weak due to its low luminosity, and in many cases they may be buried in the noise, making them difficult or even impossible to detect.

In contact binary systems, the presence of a third light source increases the overall brightness of the system, thereby affecting the photometric light curve. If significant, this effect may prevent parameters such as the mass ratio from being accurately estimated relative to its real value. Therefore, if a third-light contribution is detected in the light curve analysis, the resulting parameters should be interpreted with caution. For the system TYC 3801-1529-1, the third-light contribution was found to be $l_3/(l_1+l_2+l_3)=0.001(1)$ in our light curve solution, indicating that this value is negligible and can be ignored.

\begin{figure}
\centering
\includegraphics[width=0.5\textwidth]{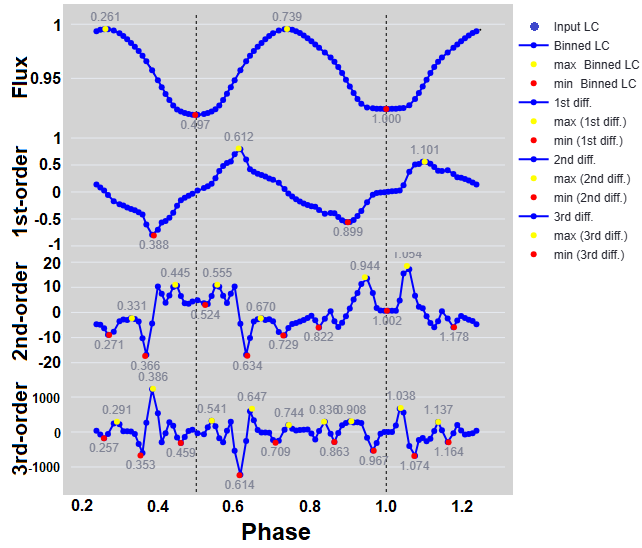}
\caption{The photometric light curve of TYC 3801-1529-1, together with its first through third derivatives (displayed from the top to the bottom panels), is presented as part of the analysis. The vertical-axis units for these panels are W m$^{-2}$, 10 W m$^{-2}$ day$^{-1}$, $10^{2}$ W m$^{-2}$ day$^{-2}$, and $10^{4}$ W m$^{-2}$ day$^{-3}$, respectively.}
\label{fig:q-Kouzuma}
\end{figure}

B) We performed two tests to investigate the range of mass ratios and compare it with our results. First: the work of \cite{2023ApJ...958...84K, 2025PASJ..tmp..108K} presented a developed technique for estimating the photometric mass ratio of overcontact binaries by examining higher-order derivatives of their light curves. This approach differs fundamentally from iterative procedures such as the widely used $q$-search or MCMC optimization, since it relies on the behavior of the second- and third-order derivatives rather than on repeated fitting. This method was also examined and discussed in the study by \cite{2024AJ....168..272P}. According to the analysis by \cite{2023ApJ...958...84K, 2025PASJ..tmp..108K} based on a large sample of synthetic light curves, distinctive patterns in these derivatives provide information that is strongly correlated with the mass ratio. After computing the third-order derivative, the times of local extrema around eclipse phases are identified and combined with the system's orbital period to obtain a parameter called $W$. Their results revealed a tight relationship between $W$ and the mass ratio $q$. Tests using systems with known spectroscopic mass ratios indicated that approximately 67\% of the photometric determinations agreed with the spectroscopic values within the estimated uncertainties, while 95\% of the cases showed deviations smaller than $\pm0.1$. The applicability of the method depends on several requirements, including the presence of clear maxima and minima in both the second- and third-order derivatives, detailed in the \cite{2023ApJ...958...84K, 2025PASJ..tmp..108K} publications. We applied the technique of \cite{2023ApJ...958...84K} to the system TYC 3801-1529-1. Based on the results, the mass ratio estimated using the method of \cite{2023ApJ...958...84K} is $q = 0.031(12)$ after some tests. This mass ratio and its associated uncertainty are consistent with those obtained from our light curve analysis as well as from \cite{2024A&A...692L...4L}. This experiment emphasizes that TYC 3801-1529-1 has the lowest known mass ratio among contact binary systems. The procedure used to estimate the mass ratio of the target system is shown in Figure \ref{fig:q-Kouzuma}.

Second: We evaluated the influence of a potential third-light contribution on the derived system parameters, particularly the mass ratio, by extending the light curve analysis beyond the assumption of a zero third-light contribution. A series of solutions was performed with the third light fixed at 0.05, 0.1, 0.2, and 0.3. The resulting fit quality and the behavior of the mass ratio and other key parameters were examined for each case. This analysis enables an assessment of whether the solution obtained under the zero third-light assumption corresponds to the global minimum in $\chi^2$ and clarifies the extent to which parameter correlations may affect the uniqueness of the solution. The $\chi^2$ value for the solution with zero third light ($l_3 = 0$) was 0.0005. When the third light was fixed at 0.05, 0.1, 0.2, and 0.3, the corresponding $\chi^2$ values increased to 0.002, 0.005, 0.01, and 0.03, respectively, while the mass ratios changed to 0.027(1), 0.029(1), 0.035(1), and 0.044(1). At $l_3 = 0.05$, only the mass ratio required adjustment, while for $l_3 = 0.1$ and 0.2 both the mass ratio and the fillout factor needed to be modified. In contrast, for fixed $l_3 = 0.3$, the remaining system parameters also changed, with deviations of approximately 5\%. These results indicate that the solution obtained under the zero third-light assumption provides the best fit in terms of $\chi^2$, while the mass ratio did not exhibit significant changes beyond the range tested in \cite{2023ApJ...958...84K}, even with increasing $l_3$.

C) The interstellar extinction ($A_V$) in the visual band for the target binary system is $A_V$=0.150(1) mag, calculated using the three-dimensional dust maps from \cite{2019ApJ...887...93G}. This result indicates that the system experiences only minor dimming and reddening due to interstellar dust along the line of sight (\citealt{2024NewA..11002227P}).

The Renormalized Unit Weight Error (RUWE, \citealt{lindegren2018re}) is extremely high for TYC 3801-1529-1. RUWE is a metric used to assess the quality of the point-source model fit in Gaia data. Values close to 1 indicate a good fit and reliable measurements, whereas values above approximately 1.4 generally signal discrepancies between the model and the observations. Such elevated values can result from the presence of multiple components in the system, contamination from nearby sources, or significant brightness variations, all of which prevent the Gaia point-source model from accurately fitting the system's photocenter. For the binary system TYC 3801-1529-1, the RUWE value in Gaia DR3 is reported as 12.08, far exceeding the range typically expected for binary systems. This high value reflects the complexity of the system and the presence of multiple components, which is also supported by the analysis of its orbital period variations in this study.

D) Based on the results in this study, we computed the ratio of spin to orbital angular momentum for TYC 3801-1529-1 using the classical relation:

\begin{equation}
\frac{J_{\rm spin}}{J_{\rm orb}} = \frac{1+q}{q} \Big[ (k_1 r_1)^2 + (k_2 r_2)^2 q \Big],
\end{equation}

\noindent where $q$ is the mass ratio, $r_1$ and $r_2$ are the fractional radii of the primary and secondary, and $k_1$ and $k_2$ are the respective gyration radii. Adopting the mass ratio and radii derived in this study, the gyration radii $k_1$ and $k_2$ were estimated from stellar structure models appropriate for the component masses and evolutionary states (\citealt{2004A&A...424..919C}), resulting in $k_1^2 = 0.06$ and $k_2^2 = 0.18$ (\citealt{poro2025}). So, we obtain $J_{\rm spin}/J_{\rm orb} \approx 1.18(15)$. This value is substantially larger than the $J_{\rm spin}/J_{\rm orb} = 0.755$ reported by \cite{2024A&A...692L...4L}, who adopted the assumption $k_1^2 = k_2^2 = 0.06$.

There are clear physical reasons why the two components of a contact binary, especially one with an extremely low mass ratio, should not be assigned the same gyration radius. The gyration radius $k$ (through $I = k^2 M R^2$) encodes the internal mass distribution of a star (\citealt{1992A&AS...96..255C, 2004A&A...424..919C, 2006epbm.book.....E}): stars with extended convective envelopes or inflated outer layers have larger $k^2$ than more centrally concentrated, radiative stars of similar mass. In low mass secondaries that are strongly Roche-distorted and possibly thermally inflated by mass transfer, the outer layers can become relatively massive compared to the core, producing a significantly larger moment of inertia. Tidal and rotational deformation in a contact configuration further redistribute mass and can increase the effective $k^2$ of the less massive component. Thus, the ad hoc choice $k_1^2 = k_2^2 = 0.06$ effectively assumes nearly identical internal structures for both stars; an assumption that is unlikely to hold in an extreme low mass contact binary and which underestimates the secondary's spin contribution.

The practical implication is important for dynamical stability. Because the secondary's $k_2$ is substantially larger here, the computed $J_{\rm spin}/J_{\rm orb}$ is driven to values well above the Darwin stability (\citealt{hut1980stability}) threshold (roughly $J_{\rm spin}/J_{\rm orb} \sim 1/3$), supporting the conclusion that the system is dynamically unstable and prone to rapid orbital evolution or merger.

E) The aliasing effect (\citealt{2012MNRAS.422.2372B}) is a well-recognized challenge in the temporal analysis of eclipsing binary systems, particularly when examining cyclic variations in O-C diagrams that are commonly interpreted as the light-time effect induced by a tertiary companion. This phenomenon was first identified in photometric timing studies of eclipsing binaries during the 1970s and 1980s, when it became clear that seasonally clustered minima could produce spurious periodicities in the derived O-C trends. Subsequent studies, including those by \cite{2006AJ....131.2986P} and \cite{2016MNRAS.455.4136B}, have emphasized the issue of annual aliasing in LITE analyses, in which the sampling pattern of the observational data itself introduces false frequency components into the period analysis.

In binary systems, when times of minima are collected in a few temporal clusters, the sampling window function acquires a characteristic frequency. If the intrinsic O-C variation exhibits a sinusoidal modulation with a period close to the sampling frequency, the combination of the two can produce an apparent period that differs from the true one. This phenomenon manifests as an artificial or shifted cyclic trend in the O-C diagram and represents the aliasing effect in period analyses.

Distinguishing between a genuine light-time effect and an alias-induced periodicity in cyclic O-C variations strongly depends on the coverage and uniformity of the observational data. When the observations span only a fraction of the modulation phase or cover a limited number of complete cycles, multiple competing periodic models may reproduce the data, potentially leading to an erroneous identification of the tertiary period. The issue becomes particularly critical when the apparent period of the tertiary companion is close to one year or its harmonics, as the sampling window frequency may interfere with the true period of the tertiary object.

\begin{figure*}
\centering
\includegraphics[width=0.99\textwidth]{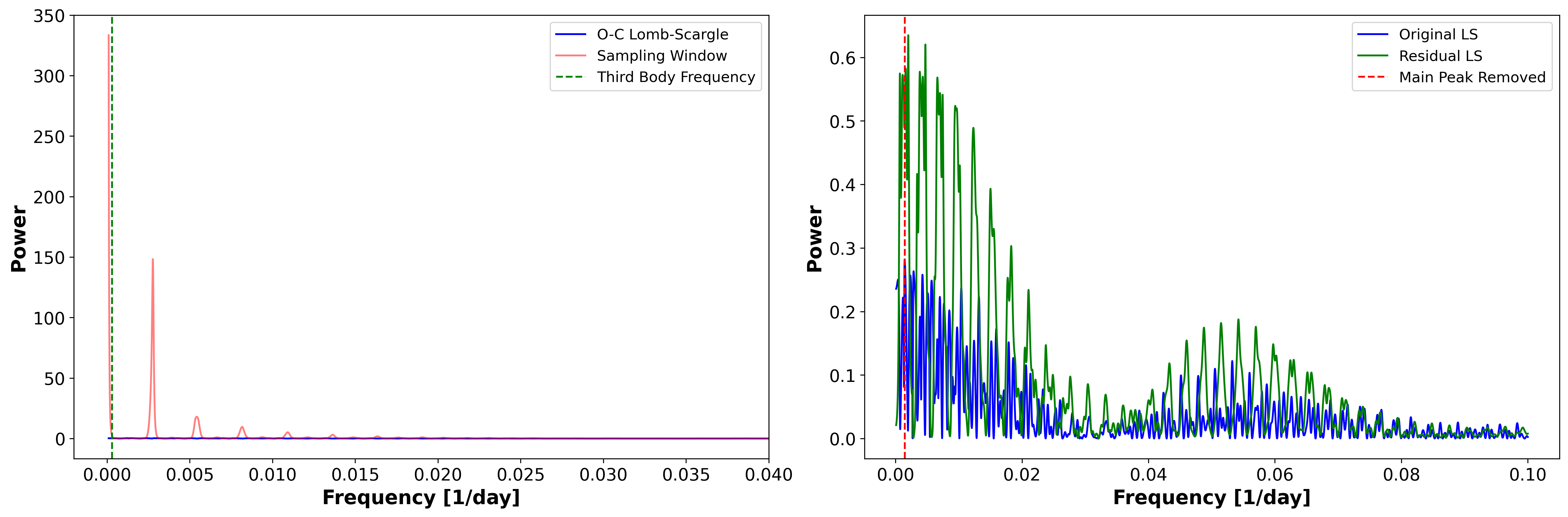}
\caption{Left: Lomb-Scargle periodogram of the O-C data along with the periodogram of the sampling window. The blue curve shows the O-C power spectrum, the red curve represents the power of the sampling window, and the green dashed line indicates the expected frequency of the third body modulation.
Right: Prewhitening test of the O-C data. The blue curve shows the original Lomb-Scargle periodogram, while the green curve represents the residuals after subtracting the sinusoidal fit at the main peak frequency (red dashed line).}
\label{fig:aliasing}
\end{figure*}

The potential impact of aliasing on the O–C data of the TYC 3801-1529-1 system was examined using the Lomb–Scargle periodogram analysis (\citealt{1982ApJ...263..835S, 2012MNRAS.422.2372B}). The extracted minima from various sources were employed to compute the frequency spectrum of the O–C values. A sampling window was also constructed based on the observation times in order to identify spurious frequencies potentially introduced by the temporal sampling pattern. The main peak in the O–C periodogram was examined, while the power spectrum of the sampling window was also inspected to assess potential aliasing (Figure \ref{fig:aliasing}left). A Monte Carlo simulation with 1000 iterations was then carried out by replacing the O–C values with random noise, yielding a $95\%$ confidence threshold for noise-induced power of $0.03664$. Because the observed peak significantly exceeds this threshold, the main O-C signal is interpreted as genuine and attributed to the third body rather than an aliasing artifact.

Moreover, a prewhitening procedure was applied to the O-C data to further examine the periodic signal attributed to the third body. The main peak in the Lomb-Scargle periodogram was first identified, and a sinusoidal model at this frequency was fitted to the data. This signal was then removed to generate residuals, and a Lomb–Scargle periodogram of the residuals was computed to check for remaining strong peaks that could indicate additional periodicities or aliasing effects (Figure \ref{fig:aliasing}right). Potential aliases corresponding to the one-year sampling frequency and its harmonics were specifically inspected. The analysis showed that after prewhitening, no significant peaks remained near the expected alias frequencies, supporting the interpretation that the observed main peak represents a genuine periodicity associated with the third body.

This analysis demonstrates that, despite the temporal clustering of the observations and the potential for aliasing, the sinusoidal signal observed in the O-C diagram of TYC 3801-1529-1 is independent of sampling effects.

F) In stellar astrophysics, the merger of contact binaries is recognized as an important evolutionary pathway that can give rise to phenomena such as luminous red novae. A well-documented example is V1309 Sco, which was initially discovered as Nova Sco 2008 (\citealt{2008IAUC.8972....1N}) but later identified as the first confirmed case of a stellar merger (\citealt{mason2010peculiar}, \citealt{tylenda2011v1309}). Its progenitor was a contact binary with an orbital period of about 1.4 days, steadily decreasing prior to the outburst (\citealt{tylenda2011v1309}). Light curve modeling indicated a very low mass ratio (\(q \approx 0.094\)), while the evolving light curve provided direct evidence of orbital decay leading to merger. Hydrodynamical simulations showed that the process was triggered by Darwin instability, which drove Roche Lobe Overflow (RLOF) and the formation of a common envelope, from which mass was ejected in a ring-like structure around the system (\citealt{nandez2014v1309}). This mass loss carried away a substantial fraction of the orbital energy and angular momentum, powering the luminous red nova event. The remnant was left as a single, differentially rotating star, and in the following years its near-infrared colors became progressively bluer, consistent with the evolution toward a blue straggler (\citealt{ivanova2013identification}, \citealt{nandez2014v1309}, \citealt{ferreira2019asymptotic}).

Given that both the ratio \(J_{\mathrm{spin}}/J_{\mathrm{orb}}\) and the mass ratio of TYC 3801-1529-1 exceed the instability threshold, this system represents a candidate for a contact binary merger. Nevertheless, \cite{2024A&A...692L...4L}, which examined the orbital period variations of this system, concluded that the period changes proceed at a stable rate over four years observations and it is increasing. Instead, the O-C analysis in this study indicates that, at least over the past six years, the system has exhibited a cyclic not a parabolic trend.

We performed an analysis to investigate the orbital period variations of the system without the effect of a third body. The sinusoidal O-C variations caused by the third-body were first fitted using a model of the form

\begin{equation}
(O-C)_\mathrm{third-body}(E) = A \, \sin\left(\frac{2 \pi E}{P_\mathrm{mod}} + \phi \right).
\end{equation}

\begin{figure*}
\centering
\includegraphics[width=0.75\textwidth]{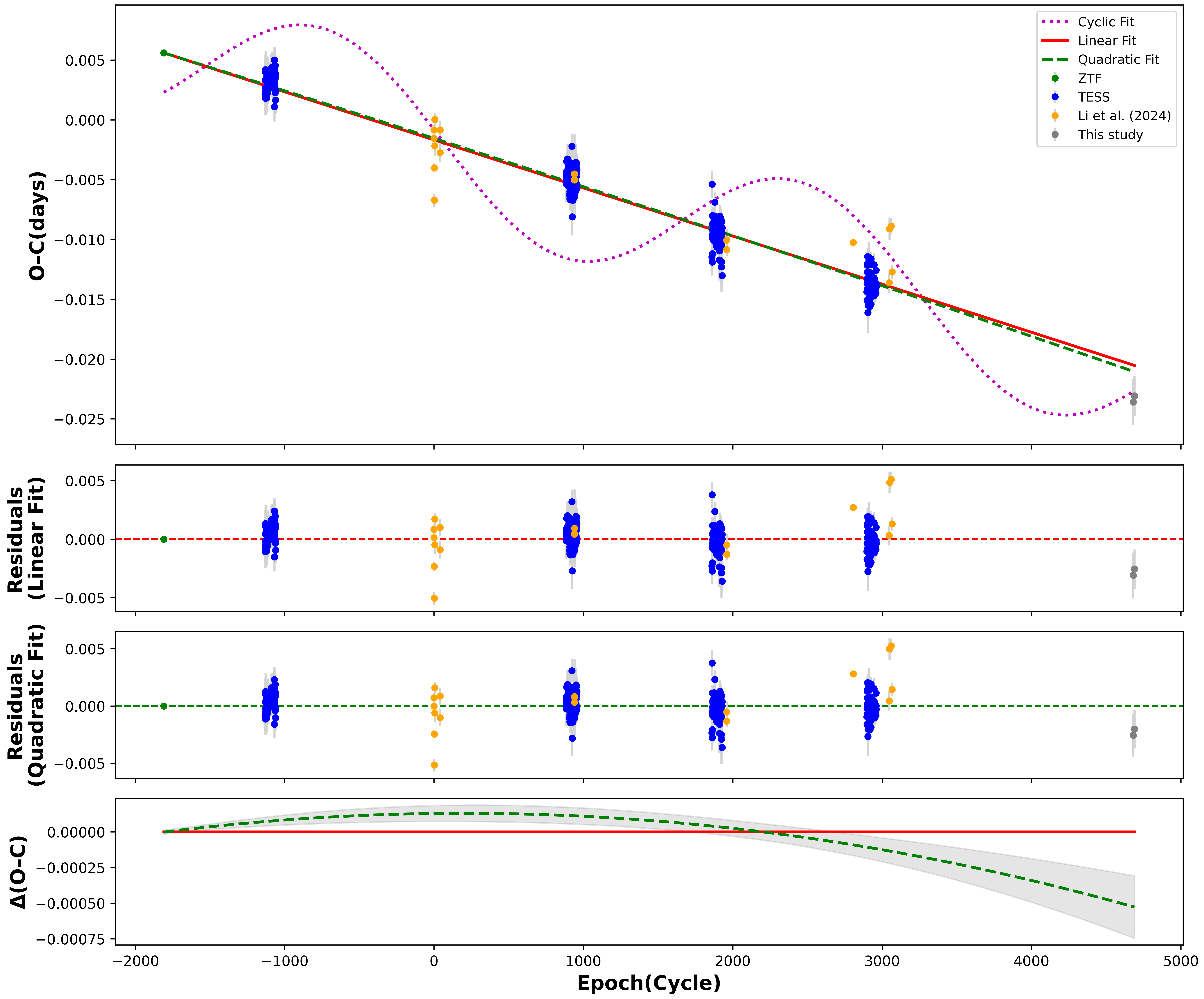}
\caption{Corrected O–C diagram of the binary system after removing the contribution of the third-body. This correction reveals the intrinsic period variations of the system. Cyclic fit comes from Figure \ref{fig:o-c}. The bottom panel shows the difference between the possible linear and parabolic fits.}
\label{fig:o-c-without-3body}
\end{figure*}

This fit allowed quantification of the light-travel time contribution of the third component. The fitted sinusoidal values were subtracted from the observed O-C at each epoch,

\begin{equation}
(O-C)_\mathrm{corr}(E) = (O-C)_\mathrm{obs}(E) - (O-C)_\mathrm{third-body}(E),
\end{equation}

\noindent producing corrected residuals that reveal the intrinsic period variations of the binary system. Linear and parabolic functions were fitted to the corrected O-C values after removing the third-body contribution to identify possible long-term trends, and the fits along with their residuals are shown in Figure \ref{fig:o-c-without-3body}. The linear fit, which shows a negative slope, gives $\chi^2 = 880$, AIC = 313, and BIC = 321, while the quadratic fit results in slightly lower $\chi^2 = 875$ and AIC = 312, but a higher BIC = 324. These close statistical values indicate that both models describe the data nearly equally well. Therefore, we adopt the simpler linear fit as the final solution. Nevertheless, if the quadratic term is physically valid, it would suggest a decreasing orbital period (Figure \ref{fig:o-c-without-3body}), which should be verified with future observations.

G) The secondary star in the TYC 3801-1529-1 system was estimated to have a mass of $M_2 = 0.05(16)\,M_\odot$ in this study. This mass of the secondary falls within the typical range of a single brown dwarf (0.013-0.08 $M_\odot$, \citealt{2014MNRAS.439.2781M}). Its temperature and radius are significantly larger than expected for such a low-mass star because it resides in a contact binary system, where mass and energy are exchanged between the components, leading to thermal and structural inflation of the secondary (\citealt{1999AJ....118.2460B}). Additionally, the primary star’s estimated mass is relatively high for its effective temperature, but comparisons with single-star models are not appropriate, as temperatures in contact binary systems are influenced by mass and energy exchange between the components. Similar behavior for the primary star has been reported in other contact binary systems, such as V899 Her (\citealt{2002A&A...387..240O}) and HN UMa (\citealt{2007ASPC..362...82O}), whose light curve solutions were constrained using spectroscopic data. The system exhibits a medium fillout factor (\citealt{2022AJ....164..202L}) and a temperature difference of 138 K between the components, placing it near thermal equilibrium. As the hotter component also possesses the greater mass, TYC 3801-1529-1 is classified as an A-subtype contact binary.

The presence of a third body in the system is indicated through the analysis of the orbital period variations. Its mass is constrained within the range $M_3 = 0.40$-$0.58\,M_\odot$, and it contributes only about 1\% of the total system light. These properties suggest that the third body is a low-mass M-type dwarf (\citealt{2024ARA&A..62..593H}). Stars of this type are typically low-luminosity, cool, and low-mass objects, which are common companions in multiple stellar systems.

H) A few close binary systems have previously been re-examined using high-resolution spectroscopy on medium- to large-aperture telescopes. For example, AW UMa was observed with the 3.6-m CFHT to reassess its contact configuration (\citealt{2015AJ....149...49R, 2025AJ....169...82R}), and AE Phe was studied using Doppler imaging with the 3.9-m AAT (\citealt{2004MNRAS.348.1321B}). In addition, adaptive optics imaging at the 3.6-m CFHT has been used to search for visual companions among close binary systems (\citealt{2007AJ....134.2353R}). High-resolution spectroscopy on large telescopes can directly constrain the mass ratio and reveal the secondary's spectral contribution, thereby providing a decisive test of the system configuration. Given the extremely low mass ratio inferred for TYC 3801-1529-1, targeted high-resolution spectroscopic follow-up would be particularly valuable to confirm the mass ratio with higher confidence and to better assess the system’s dynamical stability and merger likelihood.

\vspace{0.6cm}
\section*{Acknowledgments}
This work was carried out within the framework of the BSN project(\url{https://bsnp.info}). We used data from the European Space Agency mission Gaia(\url{http://www.cosmos.esa.int/gaia}). In this work, we utilize observations obtained by the TESS mission, supported through NASA's Explorer Program. Ground-based observations of the target system were conducted with the cooperation of the Observatorio Astron\'omico Nacional on the Sierra San Pedro M\'artir (OAN-SPM), Baja California, M\'exico. We used IRAF, distributed by the National Optical Observatories and operated by the Association of Universities for Research in Astronomy, Inc., under a cooperative agreement with the National Science Foundation. We are deeply grateful for the valuable support and constructive suggestions offered by Kaloyan Penev. We also thank Sabrina Baudart for her efforts in observing this binary system. We are grateful to Shinjirou Kouzuma for granting access to the code for mass ratio estimation using the derivative method.

\vspace{0.6cm}
\bibliography{References}{}
\bibliographystyle{aasjournal}

\end{document}